\documentstyle[12pt]{article}
\begin{document}
\noindent
\hfill{Warsaw University preprint {\it IFD/1/1994}}

\hfill{August 1994}
\vskip1cm
\begin{center}
{\large \bf LOW $Q^2$, LOW $x$ REGION IN ELECTROPRODUCTION \\
$-$ AN OVERVIEW}

\vskip0.5cm
{\large B. Bade\l{}ek$^1$} and {\large J. Kwieci\'nski$^2$}
\end{center}

\vskip0.5cm
\noindent
$^1$ {\it Department of Physics, Uppsala University, P.O.Box 530,
751 21 Uppsala, Sweden} and

\noindent
\hspace{2mm}
{\it Institute of Experimental Physics, Warsaw University, Ho\.za 69, 00-681
Warsaw, Poland}\\

\noindent
$^2$ {\it H. Niewodnicza\'nski Institute of Nuclear Physics, Radzikowskiego
152,
31-342 Cracow,}

\noindent
\hspace{2mm}
{\it Poland}

\medskip\medskip\medskip
\vskip1cm
\begin{abstract}
We summarise existing experimental and theoretical knowledge on the
structure function $F_2$ in the region of low $Q^2$ and low $x$.
The constraints on the behaviour of structure functions in the
limit of $Q^2=0$ are listed. Phenomenological low $Q^2$
parametrisations of the structure functions are collected and
their dynamical content is discussed. The high energy
photoproduction and nuclear shadowing are also briefly
described. Recent update of the low $Q^2$, low $x$ experimental data is given.
\end{abstract}

\medskip \medskip \medskip
\section{Introduction}
\medskip \medskip
Interest in the low $Q^2$, low $x$ phenomena in the inelastic lepton--hadron
scattering is connected with experimental constraints
that the low $x$ region in the present fixed target experiments can only
be reached on the expense of lowering the $Q^2$ values down to 1 GeV$^2$
or less (see e.g. ref.\cite{NMCF2}).
Unified treatment of small- and large $Q^2$ regions might also be of high
practical importance for large $Q^2$ data analysis. This stems from the fact
that in all leptoproduction experiments, including those at HERA, a radiative
corrections procedure
has to be applied in order to extract the structure functions from the data.
The radiative "tails" originating from processes at $Q^2$ values from the
interval $Q^2_{meas} \geq Q^2 \geq 0$ contribute to measurements at
$Q^2 = Q^2_{meas}$ and the knowledge of the structure functions in this
$Q^2$ interval is necessary for the iterative data unfolding procedure
\cite{radcor1,radcor2}.

\medskip
Remembering that due to the conservation of
the electromagnetic current the structure function $F_2$ must vanish in
the limit $Q^2 \rightarrow 0$, the Bjorken scaling which holds approximately
at high $Q^2$ cannot be a valid concept at low $Q^2$. Theoretical models
assuring a smooth transition from the scaling to non-scaling regions and
applicable in a wide $Q^2$ interval are thus necessary for understanding
the experimental data and the underlying dynamics. The continuity of
the physical processes occuring when passing from
low- to high $Q^2$ is illustrated in fig.1. We shall be predominantly
concerned with the region of low $Q^2$ and high $W$, i.e. beyond the
resonances.

\medskip
The need for modifying the QCD improved parton model by including
contributions to the structure functions behaving as $1/(Q^2)^n (n\geq 1)$
is clearly visible in the data \cite{nmcqcd}. These contributions, called
"higher twists" are important at moderate values of $Q^2 (\sim $1 GeV$^2$).
They
follow from the operator product expansion \cite{rgr} and describe effects of
the struck parton's interaction with target remnants thus reflecting
confinement effects. Clearly the theoretical
description unifying the confinement and deep inelastic regions should
contain terms of well defined physical origin, corresponding to such
contributions.

\medskip
The purpose of this paper is to collect the existing knowledge about the
region of low $Q^2$, low $x$ (i.e. $Q^2$ below 3 GeV$^2$ and $x$ below 0.03
or so) in a way which could at the same time make it
easy to use in practical applications. The paper should be considered
as an extension of the comprehensive review of the small $x$ physics
\cite{BCKK} towards the more detailed treatment of the
low $Q^2$ problems.  We shall be exclusively concerned with {\bf charged}
lepton inelastic scattering.  The recent review of problems specific for
inelastic neutrino (and antineutrino) interactions is presented in the ref.
\cite{KOPEL}. It is predominantly the structure function $F_2$ which will be
discussed; particular final state structures, like jets, diffractive
dissociation, etc. will not be considered.

\medskip
The content of the paper is as follows. After basic definitions and
constraints (section 2) we present
theoretical ideas and models which describe the low $Q^2$ physics (section 3).
High energy photoproduction (section 4) is then followed by a description
of phenomenological parametrisations of structure functions (section 5).
Special attention is given to dynamical models of a low $Q^2$ behaviour of
$F_2$ (section 6). Nuclear shadowing is described in section 7 and finally
an update of experimental data is given in section 8. Section 9 contains
conclusions and outlook.

\medskip \medskip \medskip
\section{Basic definitions and constraints}
\medskip \medskip
Kinematics of inelastic charged lepton scattering is defined in fig.2.  One
photon exchange approximation is assumed throughout this paper.
The imaginary part of the forward Compton scattering amplitude of the virtual
photon is defined by the tensor $W^{\mu \nu}$(see e.g.\cite{halmar}):
\begin{eqnarray}
W^{\mu \nu}(p,q)& = &{F_1(x,Q^2)\over M}\left(-g_{\mu \nu}+
{q^{\mu}q^{\nu}\over q^2}\right) + \nonumber\\
& &{F_2(x,Q^2)\over M (p\cdot q)}\left(p^{\mu}-{p\cdot q\over
q^2}q^{\mu}\right)
\left(p^{\nu}-{p\cdot q\over q^2}q^{\nu}\right)
\label{wmunu}
\end{eqnarray}
In this equation $Q^2=-q^2$ where $q^2$ is the square of the
four-momentum transfer
, $x=Q^2/(2p\cdot q)$ the Bjorken scaling variable and M is taken as
the proton mass.  The invariant quantity $p\cdot q$ is related
to the energy transfer $\nu$ in the target rest frame, $p\cdot q=M\nu$.
The invariant mass of the electroproduced hadronic system, $W$, is then
$W^2=M^2+2M\nu-Q^2$.
Often one denotes $W^2\equiv s$.

\medskip
The deep inelastic regime
is defined as a region where both $Q^2$ and $2M\nu$ are large and their
ratio, $x$, is kept fixed.
At $Q^2$ smaller than few GeV$^2$, $x$ can probably no longer be interpreted
as a momentum of a struck parton but it remains a convenient variable
for displaying the data.
The functions $F_1(x,Q^2)$ and $F_2(x,Q^2)$ are the
structure functions of the target. For a nuclear target it will be assumed
that the structure functions are normalised to the number of nucleons in the
target nucleus and denoted $F_{i}^A$, $i=$1,2 (except for the deuteron where
a symbol $F_2^d$ will be used). The tensor $W^{\mu \nu}$
satisfies the current conservation constraints:
\begin{eqnarray}
q_{\mu}W^{\mu \nu}=0 \nonumber \\
q_{\nu}W^{\mu \nu}=0
\label{cconv}
\end{eqnarray}
This follows from the fact that $W^{\mu \nu}$ is related to the matrix element
of the product of the electromagnetic current operators $j_{em}^{\mu}(x)$:
\begin{eqnarray}
W^{\mu \nu} \equiv ImT^{\mu \nu}
\propto Im~ i\int d^4z \exp(iqz)<p\mid Tj_{em}^{\mu +}(z)j_{em}^{\nu}(0)
\mid p>
\label{tprod}
\end{eqnarray}
where the symbol '$T$' denotes time ordering.

\medskip
Let us rearrange eq.(\ref{wmunu}) in order to display explicitly
the potential kinematical singularities
of the tensor $W^{\mu \nu}$ at $Q^2=0$:
\begin{eqnarray}
W^{\mu \nu}(p,q)& = &-{F_1\over M}g^{\mu \nu} +{F_2\over M(p\cdot q)}
p^{\mu}p^{\nu}
+ \nonumber\\ & &\left({F_1\over M}+{F_2\over M}
{p\cdot q\over q^2}\right){q^{\mu}q^{\nu}\over q^2} -
{F_2\over M}{p^{\mu}q^{\nu}+p^{\nu}q^{\mu}\over q^2}
\label{ksin}
\end{eqnarray}
These singularities cannot be real and appear only as artifacts of the way we
wrote up
$W^{\mu \nu}$. In order to eliminate them we have to impose the following
conditions on the structure functions $F_i$ in the limit $Q^2 \rightarrow 0$:
\begin{equation}
F_2=O(Q^2)
\label{f20}
\end{equation}
\begin{equation}
{F_1\over M}+{F_2\over M}{p\cdot q\over q^2}=O(Q^{2})
\label{f120}
\end{equation}
These conditions have to be fulfilled for arbitrary $\nu$.  They will play
important role for the parametrisations of the structure functions
at low $Q^2$.

\medskip
The differential electroproduction cross section is expressed in the following
way by the structure functions $F_i$:
\begin{eqnarray}
{d^2\sigma (x,Q^2)\over dQ^2dx}={4\pi \alpha^2\over Q^4}\left[\left(1-y-
{Mxy\over 2E}\right) {F_2(x,Q^2)\over x}
+ \left (1 - {2m^2\over Q^2}\right ) y^2F_1(x,Q^2)\right]
\label{d2sig}
\end{eqnarray}
where $E$ denotes the energy of the incident lepton in the target rest frame,
$m$ is the electron (muon) mass, $y=\nu/E$ and $\alpha$ is the electromagnetic
coupling constant.

\medskip
Instead of $F_1$ the structure function $R(x,Q^2)$ defined as
\begin{equation}
R(x,Q^2)={\sigma_L\over\sigma_T}={(1+4M^2x^2/Q^2)F_2\over 2xF_1}-1
\label{r}
\end{equation}
is often used where $\sigma_L$ and $\sigma_T$ denote the cross sections
for the longitudinally and transversally polarised virtual photons
respectively.
The differential cross section (\ref{d2sig}) then reads
\begin{eqnarray}
{d^2\sigma (x,Q^2)\over dQ^2dx}={4\pi \alpha^2\over Q^4}{F_2\over x}
\left[1-y-
{Mxy\over 2E} + \left(1 - {2m^2\over Q^2}\right )
{y^2(1+4M^2x^2/Q^2)\over 2(1+R)}\right]
\label{d2sigr}
\end{eqnarray}
\noindent
Real photons are only transversally polarised and therefore
$\sigma_L$ and $R$ vanish when $Q^2\rightarrow 0$. This vanishing follows from
eqs (\ref{f20}) and (\ref{f120}). The function $R$ is related to the
frequently used longitudinal structure function $F_L(x,Q^2)$ via $R=F_L/2xF_1$,
and
\begin{equation}
F_L(x,Q^2)=\left (1+{4M^2x^2\over Q^2}\right )F_2-2xF_1.
\label{FL}
\end{equation}
At large $Q^2$, $F_L$ is directly sensitive to the gluon distribution function,
which plays a crucial role in the interactions at small $x$, \cite{rgr}.

\medskip \medskip \medskip
\section{Basic theoretical concepts relevant for the small $Q^2$ region}
\medskip \medskip
In the leading log$Q^2$
approximation of perturbative QCD which is applicable in the high $Q^2$ region
the structure function $F_2(x,Q^2)$
is directly related to the quark- and antiquark momentum distributions,
$q_i(x,Q^2)$ and $\bar q_i(x,Q^2)$:
\begin{equation}
F_2(x,Q^2)=x\sum_i e_i^2\left[q_i(x,Q^2)+\bar q_i(x,Q^2)\right]
\label{f2pm}
\end{equation}
where '$i$' denotes quark flavours and $e_i$ are the quark charges.
At high $Q^2$ the quark and antiquark distributions
exhibit the approximate Bjorken scaling mildly violated
by the QCD logarithmic corrections. The evolution of these distributions with
$Q^2$ is described by the Altarelli-Parisi equations \cite{AP}.
These equations as well
as the relation (\ref{f2pm}) acquire corrections
proportional to $\alpha_s(Q^2)$ in the next-to-leading log$Q^2$ approximation.

\medskip
Systematic analysis of the structure functions in the Bjorken limit
can be done using the operator product expansion of the electromagnetic
currents
(cf. eq. \ref{tprod}), \cite{rgr}. This expansion leads to the expansion of the
structure functions in the inverse powers of $Q^2$:
\begin{equation}
F_2(x,Q^2)=\sum_{n=0}^{\infty}{C_n(x,Q^2)\over (Q^2)^n}
\label{ht}
\end{equation}
where the  functions $C_n(x,Q^2)$ depend weakly (i.e. logarithmically) on
$Q^2$. Various terms in this expansion are referred to as leading ($n=0$)
and higher ($n\geq 1$) twists.  The "twist" number is defined in such a way
that the leading one is equal to two and higher ones correspond to
consecutive even integers.

\medskip
Thus the right hand side of the equation (\ref{f2pm}) with approximate Bjorken
scaling of quark distributions corresponds to the  "leading twist"
contribution to the $F_2$. For moderately large values of $Q^2$ ($Q^2$ of the
order of a few GeV$^2$) contributions of the "higher twists" may become
significant. Contrary to the common opinion the higher twists are only
corrections to the leading (approximately scaling) term (\ref{f2pm}) in the
large $Q^2$  region. Thus they cannot correctly describe the low $Q^2$
(i.e. nonperturbative) region since the expansion (\ref{ht}) gives a divergent
series there. In particular it should be noted that the individual
terms in this expansion violate the constraint (\ref{f20}). In order to
correctly describe this region the (formal) expansion has to be summed
beforehand, at large $Q^2$,
and then continued to the region of $Q^2\sim 0$.  This is automatically
provided by certain models like the Vector Meson Dominance (VMD)
model. To be precise the VMD model together with its generalisation
which gives (approximate) scaling at large $Q^2$ can be represented
in a form (\ref{ht}) for sufficiently large $Q^2$.

\medskip
The VMD model is a quantitative realisation of the experimental fact
that the photon interactions are often similar to these of a hadron \cite
{bauer,gram}. The structure function $F_2$ is in this model represented by:
\begin{eqnarray}
F_{2}\left[x = Q^{2}/(s + Q^{2} - M^{2}), Q^{2}\right]& =& {Q^{2}\over 4\pi }
\sum^{}_{v}
 {M^{4}_{v}\sigma_{v}(s)\over \gamma^{2}_{v} (Q^{2} + M^{2}_{v})^{2}}
\label{vmd}
\end{eqnarray}
\noindent
The quantities $\sigma _{v}(s)$ are the
vector meson--nucleon  total
cross sections, $M_v$ is the mass of the vector meson $v$  and
$\gamma^{2}_{v}$ can  be
related in the standard way to the  leptonic  width  of  the $v$
\cite{bauer}:
\begin{equation}
{\gamma_v^2\over \pi}={\alpha^2 M_v\over 3\Gamma_{e^+e^-}}.
\label{gamv}
\end{equation}
If only the finite number of vector mesons is included in the sum (\ref{vmd})
then the $F_2$ vanishes as ${1/Q^2}$ at large $Q^2$. Therefore it does not
contain the "leading twist" term. The scaling can be introduced by including
the infinite number of vector mesons in the sum. This version of the VMD is
called the Generalised Vector Meson Dominance (GVMD) model \cite{bauer,gram}.
 The heavy mesons
contribution is directly related to the structure function in the scaling
region.

\medskip
In practical applications to the analysis of experimental data which extend to
the moderate values of $Q^2$ one often includes the higher twists corrections
in the following simplified way:
\begin{equation}
F_2(x,Q^2)=F_2^{LT}(x,Q^2)\left [1 + {H(x)\over Q^2}\right ]
\label{F2ht}
\end{equation}
where the $F_2^{LT}$ is the leading twist contribution to  $F_2$ and $H(x)$ is
determined from fit to the data. This simple minded expression may not be
justified theoretically since in principle the higher twist terms, i.e.
functions $C_n(x,Q^2)$ for $n\ge1$ in eq.(\ref{ht}) evolve differently with
$Q^2$ than the leading twist term.

\medskip
At high energies the Regge theory \cite{PCOLLINS} is often used to parametrise
the cross sections as the functions of energy $W$. The high
energy behaviour of the total cross sections is then given by the
following expression:
\begin{equation}
\sigma_t (W)=\sum_i \beta_i (W^2)^{\alpha_i-1}
\label{regge}
\end{equation}
where $\alpha_i$ are the intercepts of the Regge poles and
$\beta_i$ denote their couplings.  The intercepts $\alpha_i$ are
the universal quantities, i.e. they are independent of the
external particles or currents and depend only on the
quantum numbers of the Regge poles which are exchanged in the crossed channel.
The Regge pole exchange is to a large extent a generalisation of the particle
exchange and formally describes a pole of the partial wave
amplitude in the crossed channel in the complex angular
momentum plane.
The Regge pole corresponding
to the vacuum quantum numbers is called pomeron.  It is expected
to have the highest intercept, close to unity. It is a phenomenologically
established fact that the
energy dependence of the total hadronic and photoproduction cross sections
can be described by two
contributions: the (effective) pomeron with intercept $\alpha_P$=1.08
and the Reggeon with the intercept $\alpha_R$ close to 0.5 \cite{DLCXS}.
Since bulk of the
cross sections comes from the "soft" processes one usually refers to the
phenomenological
pomeron having its intercept slightly above unity as the "soft" pomeron.
On the other hand in the leading logarithmic approximation of
perturbative QCD  one finds
 the pomeron  with intercept
significantly above unity \cite{BFKL,GLR}.  Since perturbative QCD is only
applicable
for the description of the "hard" processes one refers to this QCD pomeron as
the "hard" pomeron.  Its relation to and interplay with the phenomenologically
determined "soft"  pomeron is still not fully understood.  It should also be
emphasised that the pomeron with intercept above unity leads to violation of
unitarity at asymptotic energies.  The unitarity is restored by multiple
scattering absorptive (or shadowing) terms.

\medskip
The Regge parametrisation of the total cross sections implies the following
parametrisation of the electroproduction structure function $F_2(x,Q^2)$:
\begin{equation}
F_2(x,Q^2)=\sum_i \beta_i(Q^2)(W^2)^{\alpha_i-1}
\label{F2rrege1}
\end{equation}
It is expected to be valid in the high energy limit and for  $W^2\gg Q^2$.
In this limit $x\simeq Q^2/W^2$ and $x\ll 1$. Therefore the Regge
parametrisation (\ref{F2rrege1}) implies the following small $x$ behaviour
of $F_2(x,Q^2)$:
\begin{equation}
F_2(x,Q^2)=\sum_i \tilde \beta_i(Q^2) x^{1-\alpha_i}
\label{F2regge2}
\end{equation}
where
\begin{equation}
\tilde \beta_i(Q^2)=(Q^2)^{\alpha_i-1}\beta_i(Q^2)
\label{beta}
\end{equation}
At low $Q^2$ the functions $\beta_i(Q^2)$ should obey the constraint
(\ref{f20}), i.e. $\beta_i(Q^2)=O(Q^2)$ for $Q^2\rightarrow 0$.

\medskip\medskip\medskip
\section{High energy photoproduction}
\medskip \medskip
The low $Q^2$ region should join smoothly the photoproduction limit, $Q^2=0$.
In this limit the following relation between the total photoproduction
cross section $\sigma_{\gamma p}(E_\gamma)$ and the structure function $F_2$
holds:
\begin{equation}
\sigma_{\gamma p}(E_{\gamma})=\lim_{Q^2\rightarrow 0}4\pi ^2\alpha
{F_2\over Q^2}
\label{siggam}
\end{equation}
This limit should be taken at fixed $\nu =E_{\gamma}$ where $E_{\gamma}$
is taken as photon energy in the laboratory frame. Several structure function
parametrisations use the photoproduction total cross section as an additional
constraint.

\medskip
When considering photoproduction it is conventional to decompose
the total cross section $\sigma_{\gamma p}$ into two parts:
\begin{equation}
\sigma_{\gamma p}=\sigma_{VMD}+\sigma_{part}
\label{sigdec}
\end{equation}

\noindent
and then

\begin{equation}
\sigma_{part}=\sigma_{direct}+\sigma_{anomalous}
\label{sigpart}
\end{equation}
In this equation $\sigma_{VMD}$ denotes the cross section corresponding
to the VMD photon--hadron interaction mechanism while $\sigma_{part}$
to the partonic mechanism
respectively. The partonic component of the cross section is next
decomposed into two terms: the "direct" term which reflects the photon
interactions with the partonic constituents of the hadron and the
"anomalous" term which corresponds to the interactions of the partons --
constituents of the photon with partonic constituents of the hadron.
In the latter case the photon
coupling to its constituents is point--like.  The main feature of the
partonic mechanism is that it corresponds to the (semi) hard
interactions which can be described by perturbative QCD.  The VMD part
on the other hand contains both hard as well as the soft components.
The former comes from the hard interaction of the partons which
are the consituents of the vector meson. Unlike the "anomalous" part this
term cannot be described by perturbative QCD alone since its
description requires knowledge of the (nonperturbative) parton distributions
in the vector meson.  The anomalous term together with the
hard part of the VMD contribution represent the point--like
interaction of the partonic constituents of the photon.  In the literature the
corresponding events are therefore called the
"resolved" photon events \cite{ACKL,levyd,ss,storrow}.

\medskip
It follows from (\ref{siggam}) that the total
photoproduction cross section can be obtained from the extrapolation
of $F_2/Q^2$ to $Q^2=0$.  The partonic component $\sigma_{part}$ (i.e.
the sum of the "direct" and the "anomalous" terms) can be identified
with the difference between the result of the extrapolation and the VMD
part \cite{ss}.  The most important distinction between various mechanisms
can be done through the analysis of the final states which will not be
discussed
here (see \cite{ss}).

\medskip
The photoproduction total cross section, like the hadron$-$hadron total
cross sections exhibits an increase with the increasing photon energy,
\cite{compil,HERAPHOT,HERAPHOTnew}, see fig.3.
This increase can be well described by the "soft" pomeron contribution
with its intercept equal to 1.08 \cite{DLCXS} (see also sec. 3).  It is also
well described by the extrapolation of the parametrisation by Abramowicz et
al.,
\cite{ALLM}.  Both predictions are shown in fig. 3.
 Another possible description of the total
cross section increase is provided by the (mini)jet production, i.e.
production of jets with relatively low $p_T$ (beginning from 1$-$2 GeV
or so). The minijet production  is described by the hard scattering
of partons coming from the photon and from the proton respectively.  The energy
dependence of the cross section reflects to a large extent the small $x$
behaviour of the parton (i.e. mostly gluon) distributions in a proton
and in a photon.  This follows from the  kinematics  of the
hard parton--parton scattering:
\begin{equation}
x_1x_2 W^2_{\gamma p} \ge 4p_T^2
\label{pt}
\end{equation}
where $x_{1,2}$ denote the momentum fractions carried by partons and
$W_{\gamma p}$ is
the total CM energy.  For increasing $W_{\gamma p}$ (and for
fixed $p_T^2$) , $x_1$ and $x_2$ may assume smaller values. The magnitude
of the cross section is sensitive to the magnitude of the minimal value
of $p_T$. It can be shown that
the multiple scattering (or absorptive) corrections are extremely important
here and they slow down significantly the increase of the total
cross section with energy \cite{ss,storrow,DURAND,forshaw}. Possible prediction
for the total cross section energy dependence which follows from the minijet
production picture are shown in fig. 3 (see \cite{HERAPHOTnew} for the
details).

\medskip
The global quantity like the total cross section is not capable to
discriminate between different models and the detailed structure
of final states may be crucial here. The relevant Monte Carlo
event generator for the minijet model with
multiple scattering is discussed in \cite{butterworth}.

\medskip \medskip \medskip
\section{Phenomenological parametrisations of structure functions}
\medskip \medskip
There exist several phenomenological parametrisations of the structure
functions which incorporate the $Q^2\rightarrow 0$ constraints (cf. sec.2)
as well as the Bjorken scaling behaviour at large $Q^2$
\cite{CHIO,brasse,NMCR,DONO,DONN,ALLM,ORSAY}. Certain
parametrisations \cite{ALLM,NMCR} also contain the (QCD motivated) scaling
violations.
The parametric form of the corresponding $Q^2$ dependence, however,
is not constrained at large $Q^2$ by the Altarelli--Parisi evolution
equations, i.e. those parametrisations are not linked with the
conventional QCD evolution.  Nor is the low $Q^2$ behaviour related
to the explicit vector meson dominance, known to dominate at low $Q^2$,
\cite{bauer}.

\medskip
{\bf A parametrisation used by CHIO Collaboration} \cite{CHIO} to fit their
data
is a combination of a simple parton model (valence quarks)
and the GVMD spirit approach (sea quarks):
\begin{equation}
F_2(x,Q^2)=P_3(2+g_3)x(1-x)^{1+g_3}+P_5{14\over 9}\,{4+g_5\over 5+g_5}(1-x)^
{1+g_5}{Q^2\over Q^2+m_0^2}
\label{chio}
\end{equation}
where $g_3=g_{03}+\varepsilon,\,\, g_5=g_{05}+\varepsilon,\,\, \varepsilon=
\kappa ln[(Q^2+m_0^2)/m_0^2]$ and $P_3, P_5, g_{03}, g_{05}, \kappa$ and $m_0$
are free parameters. This parametrisation is obsolete and has been mentioned
here for historical reasons only.

\medskip
{\bf A parametrisation by Brasse et al.} \cite{brasse} concerns the resonance
region 0.1 $\leq Q^2 \leq$ 6 GeV$^2$, 1.11 $\leq W \leq $1.99 GeV, cf. fig.1.
The virtual Compton scattering cross section is assumed as:
\begin{equation}
\sigma(Q^2,W,\epsilon)=\sigma_T(Q^2,W)+\epsilon \sigma_L(Q^2,W)
\label{brasse1}
\end{equation}
where the parameter $\epsilon$  in the formula (\ref{brasse1}) denotes the
degree of polarisation of the virtual photon and the cross sections
$\sigma_T(Q^2,W)$ and $\sigma_L(Q^2,W)$ are related in a standard way to
the structure functions $F_2$ and $F_L$:
\begin{equation}
F_2={Q^2\over 4\pi^2 \alpha}(\sigma_T+\sigma_L)
\label{brasse2}
\end{equation}
\begin{equation}
F_L={Q^2\over 4\pi^2 \alpha}\sigma_L
\label{brasse3}
\end{equation}
The cross section $\sigma(Q^2,W,\epsilon)$
 is parametrised in the following way:
\begin{equation}
ln(\sigma/G^2)=a(W) + b(W)ln{\mid\mbox{\boldmath $q$}\mid \over
\mid\mbox{\boldmath $q_0$}\mid}
 +c(W)\mid ln{\mid\mbox{\boldmath $q$}\mid \over
\mid\mbox{\boldmath$q_0$}\mid}\mid^{d(W)}
\label{brasse4}
\end{equation}
where $\mbox{\boldmath$q$}$ is the three momentum transfer to the hadronic
system , i.e. $\mid\mbox{\boldmath$q$}\mid=\sqrt{Q^2+\nu^2}$,
$\mid\mbox{\boldmath$q_0$}\mid $ is the value of
$\mid\mbox{\boldmath$q$}\mid$ for $Q^2$=0 , i.e.
 $\mid\mbox{\boldmath$q_0$}\mid=(W^2-M^2)/2M$
 and $G^2(Q^2)$ is the dipole
form factor of the nucleon, i.e.:
\begin{equation}
G^2(Q^2)=\left ({1\over 1+Q^2/0.71{\rm GeV}^2}\right)^2
\label{brasse5}
\end{equation}
The parameters of $a(W), b(W), c(W)$ and $d(W)$ were obtained from the fit to
the data in different bins of $W$.  Their values are tabulated in
ref.\cite{brasse}.

\medskip
{\bf A complete parametrisation of $F_2^d$ including the resonance region
was obtained by the NMC} \cite{NMCR}. In the resonance region $F_2^d$ was
fitted
to the data from SLAC \cite{MEST} taking only the $\Delta$(1232) resonance into
account.  Outside the resonance region a QCD based parametrisation was used
to describe the data of CHIO \cite{CHIO}, SLAC \cite{SLAC}, BCDMS
\cite{BCDMS} and EMC NA28 \cite{EMCNA28}. For this purpose,
the structure function was parametrised as:
\begin{equation}
F_2^d(x,Q^2)=[1-G^2(Q^2)][F^{dis}(x,Q^2)+F^{res}(x,Q^2)+F^{bg}(x,Q^2)]
\label{f2d}
\end{equation}
where $F^{dis}$ and $F^{res}$ are the contributions from the deep inelastic and
 resonance regions respectively and $F^{bg}$ describes the background under
the resonance.  The nucleon electromagnetic form factor is given by eq.
(\ref{brasse5}); the factor 1$-G^2$ in eq. (\ref{f2d}) suppresses $F_2$ at
low values of $Q^2$ where elastic scattering on the nucleon dominates.

\medskip
The contribution from the deep inelastic region was parametrised as
\begin{equation}
F^{dis}(x,Q^2)=\left[{5\over 18}\,{3\over B(\eta _1,\eta _2+1)}x_w^{\eta
_1}{{}}
(1-x_w)^{\eta_2}+{1\over 3}{\eta_3}(1-x_w)^{\eta_4}\right]S(x,Q^2)
\label{fdis}
\end{equation}
where $x_w=(Q^2+m_a^2)/(2m\nu +m_b^2)$ with $m_a^2$=0.351 GeV$^2$ and $m_b^2$
=1.512 GeV$^2$.  The quantity $B$ is the Euler's beta function and $\eta_1,...,
\eta_4$ are linear functions of the variable $\bar s$,
\begin{equation}
\eta_i=\alpha_i+\beta_i\bar s,  \\
\label{etai}
\end{equation}
where\\
\begin{equation}
\bar s=ln{ln[(Q^2+m_a^2)/\Lambda^2]\over ln[(Q_0^2+m_a^2)/\Lambda^2]}
\label{sbar}
\\
\end{equation}
with $Q_0^2$=2.0 GeV$^2$ and $\Lambda$=0.2 GeV. The constants $\alpha_1,....,
\alpha_4$ and $\beta_1,....,\beta_4$ were free parameters in the fit.

\medskip
The factor $S(x,Q^2)$ in eq.(\ref{fdis}) suppresses $F^{dis}$ in the
resonance region close to the single pion production threshold:
\begin{equation}
S(x,Q^2)=1-e^{-a(W-W_{thr})},
\label{sxq2}
\end{equation}
with $W_{thr}=$1.03 GeV and $a$=4.177 GeV$^{-1}$.

\medskip
The form adopted for the contribution from the resonance region was
\begin{equation}
F^{res}(x,Q^2)=\alpha_5^2G^{3/2}e^{-(W-m_{\Delta})^2/\Gamma^2},
\label{fres}
\end{equation}
with $m_\Delta$=1.232 GeV, $\Gamma$=0.0728 GeV and $\alpha_5$, a free
parameter in the fit, was equal to 0.89456.
This parametrisation takes into account only the
$\Delta$(1232) contribution; higher mass resonances are neglected.

\medskip
The background under the resonance region was parametrised as
\begin{equation}
F^{bg}(x,Q^2)=\alpha_6^2G^{1/2}\xi e^{-b(W-W_{thr})^2},
\label{fbg}
\end{equation}

where

\begin{equation}
\xi = \sqrt{{\left[(W+c)^2+M^2-m_{\pi }^2\right]^2\over 4(W+c)^2}-M^2}
\label{xi}
\nonumber
\end{equation}
with b=0.5 GeV$^{-1}$ and c=0.05 GeV.  The parameter $\alpha_6$, left free
in the fit, was equal to 0.16452.

\medskip
The parametrization (\ref{f2d}) of the function $F_2^d$ is valid from
$Q^2\sim 0$
up to about 200 GeV$^2$ and from $x = 0.003$ up to 0.7. However, the
results of the fit given in \cite{NMCR} are obsolete since the new
$F_2$ measurements by the NMC \cite{NMCF2} which were not included in the
fit differ by up to 30$\%$ from the former world data. When used later in the
radiative corrections procedure the deep inelastic part of $F_2$ was
refitted while the parametrisation of the resonances and of the
background was kept fixed. Two types of the former were used: 8-- and
15--parameter functions. The 15--parameter function and the fitted
values of its coefficients are given in Table 1 \cite{NMCF2}.

\setcounter{table}{0}
\begin{table} \label{tab1}
\begin{center}
$\displaystyle
 F_2(x, Q^2) = A(x) \cdot \left[ \frac{\ln (Q^2/\Lambda^2)}
                           {\ln (Q_0^2/\Lambda^2)} \right]^{B(x)}
                    \cdot \left[ 1 + \frac{C(x)}{Q^2} \right]; $ \\[1em]
$Q_0^2 = 20\,{\rm GeV}^2, ~~~ \Lambda = 250\,{\rm MeV}; $        \\[1em]
$A(x) = x^{a_1} (1-x)^{a_2} \left[ a_3 + a_4 (1-x) + a_5 (1-x)^2
                             + a_6 (1-x)^3 + a_7 (1-x)^4 \right];$
                                                                 \\[1em]
$B(x) = b_1 + b_2 x + b_3/(x+b_4); $ \\[1em]
$C(x) = c_1 x +c_2 x^2 + c_3 x^3 + c_4 x^4. $\\[2em]
\begin{tabular}{|c|r|r|}
\hline
  Parameter & proton     & deuteron \\
\hline
\hline
    $a_1$   & $-0.1011$  & $-0.0996$ \\
\hline
    $a_2$   & $ 2.562 $  & $ 2.489 $ \\
\hline
    $a_3$   & $ 0.4121$  & $ 0.4684$ \\
\hline
    $a_4$   & $-0.518 $  & $-1.924 $ \\
\hline
    $a_5$   & $ 5.967 $  & $ 8.159 $ \\
\hline
    $a_6$   & $-10.197$  & $-10.893$ \\
\hline
    $a_7$   & $ 4.685 $  & $ 4.535 $ \\
\hline
    $b_1$   & $ 0.364 $  & $ 0.252 $ \\
\hline
    $b_2$   & $-2.764 $  & $-2.713 $ \\
\hline
    $b_3$   & $ 0.0150$  & $ 0.0254$ \\
\hline
    $b_4$   & $ 0.0186$  & $ 0.0299$ \\
\hline
    $c_1$   & $-1.179 $  & $-1.221 $ \\
\hline
    $c_2$   & $ 8.24  $  & $ 7.50  $ \\
\hline
    $c_3$   & $-36.36 $  & $-30.49 $ \\
\hline
    $c_4$   & $ 47.76 $  & $ 40.23 $ \\
\hline
\hline
\end{tabular}
\caption{\it The parametrisation of $F_2^p$ and $F_2^d$ \protect\cite{NMCF2}.
This function is strictly valid only in the kinematic range of the NMC,
SLAC and BCDMS data.}
\end{center}
\end{table}

\medskip
{\bf A low $Q^2$ parametrisation of $F_2^p$ was obtained by Donnachie
and Landshoff} \cite{DONO}. In this parametrisation the parton model is
extrapolated down to the region of low $Q^2$ (including the photoproduction)
respecting the constraints discussed in sec.2. The parametrisations of the
sea- and valence quark distributions at small values of $x$ are based on
Regge theory (cf. sec.3)
and correspond to the pomeron intercept equal approximately 1.08 and reggeon
intercept
about 0.5. The structure function $F_2^p$ is given by the conventional parton
model formula:
\begin{equation}
F_2^p=x[{4\over 9}(u+\bar u)+{1\over 9}(d+\bar d)+{1\over 9}\lambda (s+\bar s)]
\end{equation}
where $u,\bar u$,...are the parton densities ans $\lambda$ measures the
reduced strength of the pomeron's coupling to strange quarks ($\lambda
\approx $0.6). The individual parton densities are related in the following
way to the sea- and valence quark distributions, $S(x)$ and $V(x)$
respectively.
\begin{eqnarray}
xu(x)&=&S(x)+2V(x),~~~~xd(x)=S(x)+V(x),~~~~xs(x)=S(x)\nonumber \\
x\bar u(x)&=&x\bar d(x)=x\bar s(x)=S(x)
\end{eqnarray}
The densities $S(x)$ and $V(x)$ are parametrised as follows:
\begin{eqnarray}
S(x)&=&0.17x^{-0.08}(1-x)^5,\nonumber\\
V(x)&=&1.33x^{0.56}(1-x)^3
\end{eqnarray}
The functions $S(x)$ and $V(x)$ are next multiplied by the factors $\phi _s
(Q^2)$ and $\phi_v(Q^2)$, respectively, where
\begin{eqnarray}
\phi_s(Q^2)=\left({Q^2\over Q^2+0.36}\right)^{1.08}\nonumber\\
\phi_v(Q^2)=\left({Q^2\over Q^2+0.85}\right)^{0.44}\nonumber\\
\end{eqnarray}
The powers are chosen so as to make $F_2^p$ vanish like $Q^2$ as
$Q^2\rightarrow
0$. This parametrisation has been recently extended and
modified \cite{DONN} to include in more detail the heavy quark contribution
and the region of large $x$. Instead of the scaling variable $x$ one uses:
\begin{equation}
\xi_i=x\left (1+{\mu^2_i\over Q^2}\right ) \\
\label{donxi}
\end{equation}
where the index '$i$' denotes charm, strange and light quarks. Strange and
charm distributions are as follows:
\begin{eqnarray}
xs(x,Q^2)&=&C_s{Q^2\over Q^2+a_s}\xi_s^{-\epsilon}(1-\xi_s)^7 \nonumber \\
xc(x,Q^2)&=&C_c{Q^2\over Q^2+a_c}\xi_c^{-\epsilon}(1-\xi_c)^7
\label{donh}
\end{eqnarray}
where $C_s \approx$ 0.22, $C_c=$0.032, $a_s=$1 GeV$^2$, $a_c=$ 6.25 GeV$^2$ and
$\epsilon=$ 0.0808. Parameters $\mu_{s,c}^2$ defining the variables $\xi_{s,c}$
(cf.~ eq.~ (\ref{donxi})) are: $\mu_s^2=$1.7 GeV$^2$ and $\mu_c^2=$16 GeV$^2$.

\medskip
Parametrisation of the light quark distributions is more involved, i.e.
different functional forms obtained from a fit to the data are used for $\xi$
smaller and greater than a certain $\xi_0$ which is a free parameter in the
fit. The valence quark distributions are parametrised as below:
\begin{eqnarray}
xu_v(x,Q^2)&=&U(\xi)\psi(Q^2) \nonumber \\
U(\xi)&=&B_u \xi ^{0.4525}~~~~~~~~~~~~~~\xi  < \xi_0
\end{eqnarray}
and
\begin{eqnarray}
xd_v(x,Q^2)&=&D(\xi)\psi(Q^2) \nonumber \\
D(\xi)&=&B_d \xi^{0.4525}~~~~~~~~~~~~~\xi <\xi_0 \nonumber \\
D(\xi)&=&\beta_d \xi^{\lambda_d}(1-\xi)^4~~~~~~~\xi >\xi_0
\end{eqnarray}
The parameter $\mu ^2$ defining the variable $\xi$ for light quarks
is $\mu^2=$0.28 GeV$^2$. Parameters $B_{u,d}$, $\beta_{u,d}$ and
$\lambda_{u,d}$
are fixed by requiring that the two functional forms defining $U(\xi)$ and
$D(\xi)$ join smoothly at $\xi_0$  and by imposing the number sum rules.
The function $\psi$ is $\psi(Q^2)=Q^2/(Q^2+b)$. The light sea quark
distributions are:

\begin{eqnarray}
q_S(x,Q^2)&=&C\xi^{-0.0808}\Phi(Q^2)~~~~~~~~~~\xi <\xi_0 \nonumber \\
q_S(x,Q^2)&=&\gamma\xi^{\lambda_s}(1-\xi)^7\Phi(Q^2)~~~~~~\xi >\xi_0
\end{eqnarray}
where $\Phi(Q^2)=Q^2/(Q^2+a)$. The parameters $C, a$ and $b$ are fixed
by fitting to the data on photo- and electroproduction. The model contains also
a Regge-like term in the sea quark distributions which behaves like
$\xi^{0.4525}$ at small $x$. It also contains an extra "higher twist" term
in the $F_2$ which is parametrised as below:
\begin{equation}
ht(x,Q^2)=D{x^2(1-\xi)^2\over   1+Q^2/Q^2_0}
\end{equation}
The values of other parameters defining the model are given in \cite{DONN}.
The parametrisation is valid for 0$ < Q^2 < $10 GeV$^2$ and for $x \geq $0.008.

\medskip
The function $F_2^p(x,Q^2)$ resulting from the parametrisation of Donnachie
and Landshoff is shown in fig.4 together with the data.

\medskip
{\bf A parametrisation proposed by Abramowicz et al.}, \cite{ALLM}, is a result
of a fit to the following structure function data sets: SLAC \cite{SLAC}, BCDMS
\cite {BCDMS} and
EMC NA28 \cite{EMCNA28} and the then available photoproduction cross section.
 The fitted function
is based on the parton model with the QCD motivated scaling violation
appropriately extrapolated to the region of low $Q^2$ (including the
photoproduction). The Regge ideas are used in the parametrisation.

\medskip
The structure function $F_2$ is decomposed into two terms, $F_2^{{\it P}}$
and $F_2^{{\it R}}$, corresponding to pomeron and reggeon exchange,
\begin{equation}
F_2=F_2^{{\it P}}+F_2^{{\it R}}
\label{af2}
\end{equation}
with each of them expressed as follows:
\begin{equation}
F_2^r={Q^2\over Q^2+m_0^2}C_r(t)x_r^{a_r(t)}(1-x)^{b_r(t)}
\label{af2r}
\end{equation}
where
\begin{equation}
{1\over x_r}={2M\nu+m_r^2\over Q^2+m_r^2}
\label{axr}
\end{equation}
and $r=P,R$ stands either for pomeron or for reggeon. The argument of the
functions $C_r(t),a_r(t)$ and $b_r(t)$ is defined as:
\begin{equation}
t=ln\left({ln[(Q^2+Q_0^2)/\Lambda^2]\over ln(Q_0^2/\Lambda^2)}\right)
\label{AT}
\end{equation}
The functions $C_{{\it R}},b_{{\it R}},a_{{\it R}}$ and $b_{{\it P}}$
which increase with $Q^2$ were assumed to be of the form:
\begin{equation}
p(t)=p_1+(p_1-p_2)t^{p_3}
\label{AP}
\end{equation}
while the functions $C_{{\it P}}$ and $a_{{\it P}}$ were assumed in the form:
\begin{equation}
p(t)=p_1+(p_1-p_2)\left ({1\over 1+t^{p_3}}-1\right)
\label{APP}
\end{equation}
The parameters $m_0^2, m_r^2, Q_0^2$ as well as the parameters $a_i,b_i$
and $C_i$ (separately  for the pomeron and the reggeon terms, $i$=1,2,3)
were obtained from the fit to the data.

\medskip
The fit of Abramowicz et al. is valid from $Q^2=$0 up to the highest $Q^2$
values obtained in the fixed target experiments, excluding the resonance
region,
$W<$1.75 GeV. The fit given in \cite{ALLM} should, however, be updated
since the recent NMC data on $F_2$ \cite{NMCF2} were not included.
The structure function $F_2^p(x,Q^2)$ which follows from this representation
is shown in fig.5.

\medskip
{\bf A parametrisation by Capella et al.} \cite{ORSAY} is based on a fit
to the fixed target data as well as to the most recent large $Q^2$ data
from HERA, \cite{H1,ZEUS} which show the significant increase of $F_2$
with decreasing $x$. It also uses the data on the photoproduction
cross section
including the highest energy results from HERA \cite{HERAPHOT,HERAPHOTnew}.
This parametrisation assumes a Regge form of $F_2$
at small $x$ and counting rules at large $x$.  The leading small $x$ behaviour
corresponding to
the pomeron is parametrised by the  $Q^2$ dependent effective pomeron intercept
1+$\Delta(Q^2)$ which interpolates between the "soft" pomeron at $Q^2$=0 and
the "hard" one at large $Q^2$.  The unusual $Q^2$ dependence of the
effective pomeron intercept is meant to describe in a phenomenological way
the absorptive corrections to $F_2$ which are expected to be strong at low
$Q^2$ values. The small $x$ behaviour corresponding to the
Reggeon contribution is parametrised in a conventional way, cf.eqs
(\ref{F2regge2},\ref{beta}).  To be precise
the structure function $F_2(x,Q^2)$ is parametrised in the following form:
\begin{eqnarray}
F_2(x,Q^2)&=&Ax^{-\Delta(Q^2)}(1-x)^{n(Q^2)+4}\left({Q^2\over Q^2+a}\right
 )^{1+\Delta(Q^2)} \nonumber \\
&+&Bx^{1-\alpha_R}(1-x)^{n(Q^2)}\left({Q^2\over Q^2+b}\right )^{\alpha_R}
\label{orsayf2}
\end{eqnarray}
The first term in (\ref{orsayf2}) corresponds to the
pomeron contribution where the effective
pomeron intercept is parametrised as:
\begin{equation}
\Delta(Q^2)=\Delta_0\left(1+{2Q^2\over Q^2+d}\right)
\label{DELTA}
\end{equation}
with $\Delta_0 \sim 0.08$ that corresponds to the intercept of the "soft"
pomeron.  The second term in (\ref{orsayf2}) corresponds to the reggeon
contribution and at large $Q^2$ can be identified with the valence quarks.
The pomeron part corresponds to the sea quark contribution at large $Q^2$.
The function $n(Q^2)$ is parametrised as follows:
\begin{equation}
n(Q^2)={3\over 2}\left(1+{Q^2\over Q^2+c}\right)
\label{NQ2}
\end{equation}
Parameters $A, B, a, b, c, d$ are obtained from the fit to the data.
The parametrisation (\ref{orsayf2}) is used for moderately small and small
$Q^2$ (including
photoproduction, i.e. $Q^2$=0).  The region of large $Q^2$ is described by
the combination of this parametrisation with the Altarelli--Parisi evolution
equations.

\medskip \medskip \medskip
\section{Dynamical models of the low $Q^2$ behaviour of $F_2$}
\medskip \medskip
In the previous section we have presented various parametrisations
of the structure functions.  Those parametrisations although being motivated
dynamically were not linked with conventional QCD evolution at large $Q^2$
(except of that by Capella et al., \cite{ORSAY}).
Nor was the low $Q^2$ dynamics explicitely taken into account.
In this section we present parametrisations
which will contain the QCD evolution. One of them will also
include the VMD dynamics, dominating at low $Q^2$.

\medskip
{\bf Dynamically calculated structure functions by Gl\"{u}ck, Reya and Vogt
} \cite{GRV1,GRV2,GRV3} extend the QCD evolution equations down to the very
low $Q^2$ region
($Q^2 <$  1 GeV$^2$). The parton distributions are calculated by evolving
from the input valence-like distributions provided at the very low scale
$\mu^2=$0.3 GeV$^2$ and both the leading and next-to-leading order
approximations are used.
The valence-like shape at the scale $\mu^2$ is assumed
not only for the valence quarks but also for gluon and sea quark
distributions:
\begin{equation}
xg(x, \mu^2)=Ax^{\alpha}(1-x)^{\beta},~~~~~x\bar q(x,\mu^2)=A'x^{\alpha '}
(1-x)^{\beta '}
\label{gluckquark}
\end{equation}
The parametrisation of the valence quarks is provided at the large scale
$Q^2=$10 GeV$^2$ and its form at $Q^2=\mu^2$ is obtained from the backward
QCD evolution. Parameters describing the parton distributions come from
a fit to the data at large $Q^2$ and are also constrained by the sum rules.
As a result of the QCD evolution the gluon and sea quark distributions
(multiplied by $x$) become immediately singular at small $x$ for $Q^2>\mu^2$.
In the leading twist approximation the absolute value of the slope in $x$
of these distributions and so the absolute value of the slope in $x$
of the $F_2$ grows with increasing $Q^2$.
Naturally the model which is based on perturbative
QCD cannot be extended into the region of very low $Q^2 \leq \Lambda^2$
where $\Lambda$ is the QCD scale parameter since the
(perturbatively calculated)
QCD coupling $\alpha_s(Q^2) \rightarrow \infty$ for $Q^2 \rightarrow
\Lambda^2$. Moreover the measurable quantities like the $F_2$ may contain
non--negligible higher twist contribution in the region of moderately
small $Q^2$, cf. eq. (\ref{ht}). Presence of higher twists in an observable
should not affect
the QCD evolution of the leading twist parton distribution since partonic
distributions of different twists evolve separately.
Surprisingly the QCD evolution turns out to be stable and gives
positive definite parton distributions down
to the very small scale $Q^2 \cong (2\Lambda)^2$.
Also it turns out that in the next-to-leading order approximation of
perturbative QCD, the $F_2$ is more stable than the distributions of partons
\cite{GRV3}.
Although the higher twist contribution can in principle be important in the
low $Q^2$ region, the leading twist $F_2$ calculated within the model
has recently
been succesfully confronted \cite{GRV2} with the low $x$ and moderate $Q^2$
($Q^2$ equal few GeV$^2$) NMC data.
The high $Q^2$, low $x$ results of HERA are resonably described
too.

\medskip
The explicit forms of the parton distributions' parametrisations are as
follows:
\begin{equation}
xv(x,Q^2)=Nx^{a}(1+A\sqrt{x}+Bx)(1-x)^D
\nonumber
\end{equation}

\noindent
for valence quarks

\begin{equation}
xw(x,Q^2)=\left [x^a(A+Bx+Cx^2)\left (ln{1\over x}\right)^b+s^{\alpha}exp\left
(-E+\sqrt{E's^\beta ln{1\over x}}\right)\right](1-x)^D
\label{gluckparf}
\end{equation}

\noindent
for gluon and light sea quarks and
\begin{equation}
xw'(x,Q^2)={(s-s_{w'})^{\alpha}\over \left(ln{1\over x}\right)^a}(1+A\sqrt{x}+
Bx)(1-x)^D~exp\left(-E+\sqrt{E's^{\beta }ln{1\over  x}}\right)
\label{gluckpar}
\end{equation}

\noindent
for heavy sea quarks, where $s$ in the formulae
(\ref{gluckparf}--\ref{gluckpar}) is:
\begin{equation}
s=ln{ln[Q^2/\Lambda^2]\over ln[\mu^2/\Lambda^2]}.
\label{s}
\end{equation}
In these equations $A,B,C,D,E,E',a$ and $b$ are low order polynomials
of $s$. These polynomials as well as values of remaining parameters can be
found in \cite{GRV1}.

\medskip
{\bf Electroproduction structure function $F_2$ in the low $Q^2$,
low $x$ region by Bade\l ek and Kwieci\'nski} \cite
{BKFZ,BKFPL}.  Contributions from both the parton model with QCD corrections
suitably extended to the low $Q^2$ region and from the low mass vector mesons
were taken into account.  The former contributions results from the large
$Q^2$ structure function analysis \cite{MRS,MRSR} which includes the recent
$F_2$ measurements by the NMC \cite{NMCF2}.

\medskip
The starting point is the Generalised Vector Meson Dominance (GVMD)
representation  of  the  structure  function
$F_{2}(x,Q^{2})$ \cite{BKFZ}:
\begin{eqnarray}
F_{2}[x = Q^{2}/(s + Q^{2} - M^{2}),Q^{2}]&=& {Q^{2}\over 4\pi } \sum^{}_{v}
 {M^{4}_{v}\sigma_{v}(s)\over \gamma^{2}_{v} (Q^{2} + M^{2}_{v})^{2}}
     +Q^{2}
 \int^{\infty }_{Q^2_0} dQ'^{2} {\Phi (Q'^{2}, s) \over (Q'^{2} + Q^{2})^{2}}
\nonumber \\
   &  \equiv &  F^{(v)}_{2}(x, Q^{2}) + F^{(p)}_{2}(x, Q^{2})\
\label{GVMDF2}
\end{eqnarray}
\noindent
The function $\Phi (Q^{2},s)$ is expressed as follows:
\begin{equation}
\Phi (Q^{2},s) = - {1\over \pi } Im \int^{-Q^2} {dQ'^2 \over Q'^2} F^{AS}_2
(x', Q'^{2})
\end{equation}
\noindent The asymptotic structure function $F^{AS}_{2}(x, Q^{2})$ is
 assumed to be given. It may be obtained from the QCD structure function
analysis in the large $Q^2$ region.
By construction, $F_{2}(x,Q^{2}) \rightarrow F^{AS}_{2}(x, Q^{2})$ for large
 $Q^{2}$.  The second term in  (\ref{GVMDF2})  can
be looked upon as the extrapolation of the (QCD  improved)  parton  model  for
arbitrary $Q^{2}$.  The first  term  corresponds  to  the  low  mass
  vector  meson
dominance part since  the  sum  extends  over  the  low  mass  vector  mesons.
Contribution of vector mesons heavier then $Q_{0}$ is included in the
 integral  in
(\ref{GVMDF2}).  Choosing the parameter $Q^{2}_{0} > (M^{2}_{v})_{max}$
where $(M_{v})_{max}$ is  the  mass  of  the
heaviest vector meson  included  in  the  sum  one  explicitly  avoids  double
counting when adding two separate contributions  to  the  structure  function.
Note that $Q_{0}$ should be smaller than the mass of the lightest
vector meson  not
included in the sum. The representation (\ref{GVMDF2})
is written for fixed $s$
and is expected to be valid at $s \gg Q^{2}$, i.e. at low $x$ but for
 arbitrary $Q^{2}$.

\medskip
In \cite{BKFPL} the representation (\ref{GVMDF2}) for  the  partonic  part
$F^{(p)}_{2}(x, Q^{2})$ was simplified as follows:
\begin{equation}
F^{(p)}_{2} (x, Q^{2}) = {Q^{2}\over (Q^{2} + Q^{2}_{0})} F^{AS}_{2}(\bar{x},
Q^{2} + Q^{2}_{0})
\label{PARTF2}
\end{equation}

\noindent where

\begin{equation}
\bar{x} = {Q^{2} + Q^{2}_{0}\over s + Q^{2} - M^{2} + Q^{2}_{0}} \equiv
{Q^{2} + Q^{2}_{0}\over 2M\nu + Q^{2}_{0}}
\label{xbar}
\end{equation}

\noindent Simplified parametrisation (\ref{PARTF2}) connecting
$F^{(p)}_{2}(x,Q^{2})$ to
$F^{AS}_{2}$ by an  appropriate
change of the arguments of the latter posseses all the main properties of  the
second term in (\ref{GVMDF2}).  First of all it is evident
that $F^{(p)}_{2}(x,Q^{2})
\rightarrow  F^{AS}_{2}(x, Q^{2})$
for large $Q^{2}$.  Moreover the parametrisation of $F^{(p)}_{2}$ defined  by
(\ref{PARTF2})  preserves
the analytic properties of the second term in the eq.(\ref{GVMDF2}).

\medskip
It should  be  stressed  that  apart  from  the  parameter $Q^{2}_{0}$  which
is constrained by physical requirements described above  the  representation
(\ref{GVMDF2})
does not contain any other free parameters except of course  those  which  are
implicitly present in the parametrisation  of  parton  distributions  defining
$F^{AS}_{2}$.

\medskip
The  proton  structure  function $F_{2}^p$  calculated  from  the
representation (\ref{GVMDF2}) with $F^{(p)}_{2}(x, Q^{2})$ given by the
eq.(\ref{PARTF2}) is shown in fig.6.
In an "approach to scaling" (i.e. in the $Q^{2}$ dependence of $F_{2}$ at
 $Q^{2}$ less  than  1~ GeV$^{2}$
or so) visible in fig.6 the change of curvature comes  from  the
factor $Q^{2}$ in eq. (\ref{GVMDF2}) and  the  magnitude  of  this  change,
particularily at
$Q^{2} \ll $ 1 GeV$^{2}$, is controlled by the vector meson contribution, a
 non--trivial  test of this mechanism. Although the vector meson contribution
$F^{(v)}_2$ dominates in the very low $Q^2$ region ($Q^2 <$ 1~ GeV$^2$) the
partonic component $F^{(p)}_2$ still gives a significant contribution (i.e. at
least 20$\%$ -- 30$\%$) there.
 At $Q^{2} =$ 10 GeV$^{2}$ the  structure  function $F_{2}$
 calculated from  the  model
differs from $F^{AS}_{2}(x, Q^{2})$ by less than $3\%$ .

\medskip
 The $x$ dependence at fixed $Q^{2}$ reflects both the
energy dependence of $\sigma _{v}$ and
the $x$-dependence of $F^{AS}_{2}$.
Expectations  coming
from Regge theory are incorporated in the parametrisations of  the
$\sigma_{v}(s)$, \cite{BKFZ}
and of the input parton distributions  at  the  reference  scale $Q^{2} = $
4 GeV$^{2}$, \cite{MRS}.
$F_2(x,Q^2)$ increases with  decreasing $x$ and that reflects
the increase of the total cross sections $\sigma_v(s)$ with increasing $s$ as
well as the increase  of $F_2^{(p)}(x,Q^2)$ with decreasing $x$. The increase
of
 $F_2^{(p)}(x,Q^2)$ with decreasing $x$ is weaker at small $Q^2$ than the
similar increase in the large $Q^2$ region. The $s$
dependence of $\sigma_v(s)$ is also relatively weak in the relevant region of
$s$, \cite{BKFZ}. As the result the $x$ dependence of $F_2(x,Q^2)$ is
relatively weak for low $Q^2$.

\medskip
The model is applicable in the small $x$ region ($x<$0.1) and for arbitrary
value of $Q^2$, however the way of parametrising the total vector
meson--nucleon cross sections imposes additional constraint: $\nu >$10 GeV.

\medskip
{\bf A parametrisation by Schuler and Sj\"ostrand} \cite{ss} extends those by
Abramowicz et al. and by Donnachie and Landshoff
including explicitly the damping of  structure functions at small $x$ which is
implied by screening effects in QCD \cite{GLR,LR1} and taking into account
the logarithmic scaling violations at large $Q^2$.  Extrapolations towards the
region of small $Q^2$ are done applying the same damping factors as in
refs. \cite{ALLM,DONO,DONN}.  In the region of low $x$ the Regge
parametrisation
is used.  In practice the $(x,\mu^2)$ plane ($\mu^2=Q^2$)
is divided into four regions as shown in fig.7
and  different parametrisation of parton distributions is assumed separately in
each region. For large $\mu^2$ ($\mu^2>\mu_0^2$) the regions of large
 and small
$x$ (regions I and III in fig.7) are divided by the boundary curve $\mu_B^2(x)$
\begin{equation}
\mu_B^2(x)=2+0.053^2 exp\left(3.56 \sqrt{ln{1\over 3x}}\right).
\label{bcurve}
\end{equation}
In the region I of large $\mu^2$ and large $x$
the structure functions
$F_2$ is given by the QCD improved parton model with the scale dependent
quark distributions $f_{v,s}(x,\mu^2)$ where the
subscripts
$v$ and $s$ denote the valence and sea distributions. The
distributions $f_{v,s}(x,\mu^2)$ satisfy the Altarelli-Parisi
evolution equations.  In the regions II--IV the structure
function $F_2$ is again expressed in terms of the quark
distributions $\hat f_{v,s}(x,\mu^2)$ with their $\mu^2$ dependence
being different in different regions separately.
In the region II of small $\mu^2$ and large $x$ ($x>x_0$) the structure
functions are damped by the $\mu^2$ dependent factors
and valence distributions are made harder, i.e.
\begin{eqnarray}
\hat f_v(x,\mu^2)&=& \left( {\mu^2\over \mu_0^2} {\mu_0^2+m_R^2\over
\mu^2+m_R^2}\right)^{(1-\eta)(1-x)/(1-x_0)}f_v(x,\mu_0^2)
\label{fv2} \\
\hat f_s(x,\mu^2)&=& \left( {\mu^2\over \mu_0^2} {\mu_0^2+m_P^2\over
\mu^2+m_P^2}\right)^{1+\epsilon}f_s(x,\mu_0^2)
\label{fs2}
\end{eqnarray}
In the region III of large $\mu^2$($\mu_0^2<\mu^2<\mu_B^2(x)$)
and small $x$, the sea and valence quark distributions are
parametrised as below:
\begin{equation}
x\hat f_v(x,\mu^2)=N_1\left({x\over x_0}\right)^{\eta}x_0f_v(x_0,\mu_0^2)+
N_2xf_v(x,\mu_B^2(x))
\label{fv3}
\end{equation}
\begin{equation}
x\hat f_s(x,\mu^2)=N_1\left({x\over x_0}\right)^{-\epsilon}x_0f_s(x_0,\mu_0^2)+
N_2xf_s(x,\mu_B^2(x))
\label{fs3}
\end{equation}
where
\begin{equation}
N_1={ln(\mu_B^2/\mu^2)\over ln(\mu_B^2/\mu_0^2)}
\label{N1}
\end{equation}
\begin{equation}
N_2=1-N_1
\label{N2}
\end{equation}
Finally in the region IV of small $x$ and small $\mu^2$ the
valence and sea quark distributions are parametrised in the
Regge form with the appropriate damping factors which guarantee
vanishing of the structure function $F_2$ in the limit $\mu^2=0$
\begin{equation}
x\hat
f_v(x,\mu^2)=\left({x\over x_0}\right)^{\eta}
\left[N_1\left({\mu_0^2+\mu_R^2\over
\mu_0^2}\right)^{1-\eta}x_0f_v(x_0,\mu_0^2)+N_2x_0^{\eta}N^v\right]
\label{fv4}
\end{equation}
\begin{equation}
x\hat f_s(x,\mu^2)=\left({x\over x_0}\right)^{-\epsilon}\left[N_1\left(
{\mu_0^2+\mu_R^2\over
\mu_0^2}\right)^{1+\epsilon}x_0f_s(x_0,\mu_0^2)+N_2x_0^{-\epsilon}N^s\right]
\label{fs4}
\end{equation}
where now
\begin{equation}
N_1=1-{x_0-x\over x_0} {\mu_0^2-\mu^2\over \mu_0^2}
\label{N11}
\end{equation}
\begin{equation}
N_2=1-N_1
\label{N22}
\end{equation}
The parameters $\epsilon$ and $\eta$ correspond to the (effective) intercepts
of the Pomeron and the Reggeon respectively ($\epsilon=0.56$, $\eta=0.45$).
The
values of the remaining parameters appearing in the formulae (\ref{fv2}) -
(\ref{N11}) are as follows: $\mu_0^2=5$ GeV$^2$, $x_0=0.0069, m_R^2=0.92$
GeV$^2,
m_P^2=0.38$ GeV$^2, N_d^v=0.121, N_u^v=2N_d^v, N_d^s=N_u^s=N_{\bar d}^s=
N_{\bar u}^s=0.044$ and $N_s^s=N_{\bar s}^s=0.5N_d^s$.  The structure function
$F_2(x,Q^2)$ is expressed in terms of the quark (and antiquark) effective
distribution functions $\hat f_i(x,Q^2)$  using the parton model formula:
\begin{equation}
F_2(x,Q^2)=x\sum_i e_i^2\hat f_i(x,Q^2).
\label{ssf2}
\end{equation}

\medskip
\noindent
Comparison of different structure function parametrisations is shown in
fig.8.

\medskip \medskip \medskip
\section{Nuclear shadowing}
\medskip \medskip
Nuclear effects in the deep inelastic lepton scattering may be experimentally
inferred from inspecting the ratio $F_2^A/F_2^d$. In this method one tacitly
assumes that the nuclear effects in $F_2^d$ can be neglected or that
$F_2^d=F_2^N$ where $F_2^N$ describes a free nucleon. In particular the nuclear
shadowing corresponds to the $F_2^A/F_2^d$ ratio being smaller than unity at
small $x$. In this sense shadowing can be regarded as a part of the so called
EMC effect as illustrated in fig.9.

\medskip
The nuclear shadowing is a firmly established experimental fact, cf.sec.8,
which has been observed both in the low $Q^2$
(including the photoproduction, i.e. $Q^2=0$) and in the large $Q^2$
region. The nuclear structure function $F_{2}^{A}$ is then:
\begin {equation}
F_{2}^{A}=F_{2}^{N}-\delta F_{2}^{A}
\end{equation}
where $\delta F_{2}^{A}$ denotes the shadowing term ($\delta F_{2}^{A}>0$).

\medskip
In the region of low $Q^2$ (and for photoproduction) the natural
and presumably the dominant mechanism of nuclear shadowing is the multiple
scattering of vector mesons in the nucleus, fig.10. The vector mesons couple
to virtual (or real) photons.
This model gives the following contribution of the shadowing to the nuclear
structure function:
\begin{equation}
\delta F_{2}^{A}={Q^2\over 4 \pi} \sum_v {M_v^4\delta \sigma_{v}^{A}\over
\gamma_v^2(Q^2+M_v^2)^2}
\label{delta}
\end {equation}
where the cross section $\delta \sigma_{v}^{A}$ is that
part of the vector meson -- nucleus total  cross section $\sigma_{v}^{A}$
(normalised to a nucleon) which corresponds to multiple scattering i.e.:
\begin{equation}
       \sigma_{v}^{A}=\sigma_{v}^{N}-\delta \sigma_{v}^{A}.
\label{SIGSHAD}
\end{equation}
Here $\sigma_{v}^{N}$  is the vector meson -- nucleon total cross section and
the remaining quantities in the formula (\ref{delta}) are defined
in sec.3. The cross section which corresponds to the multiple scattering can be
obtained from the Glauber theory \cite{GLAUBER,bauer}.
The negative sign in eq.(\ref{SIGSHAD}) reflects the fact that
the vector meson--nucleon scattering amplitude is assumed to be imaginary.
Thus the shadowing in the inelastic lepton--nucleus scattering, in the VMD
reflects the absorptive character
of the elementary vector meson--nucleon scattering amplitude.
It follows from eq.(\ref{delta}) that the shadowing term which corresponds
to the multiple rescattering of (finite number of) vector mesons vanishes
for large $Q^2$.

\medskip
     At large $Q^2$ one expects the parton model to be applicable. It leads
to the Bjorken scaling mildly violated by the perturbative QCD corrections.
The parton model is described by the "hand--bag" diagram of fig.11.
It is this "hand--bag" structure and the point-like coupling
of the photons to  partons (i.e. to quarks and antiquarks)  which guarantees
the Bjorken scaling (modulo the
perturbative QCD corrections) independently of the structure of the lower
part of the diagram. The nuclear shadowing in the large $Q^2$ region may
come from the multiple interaction  contributions to the lower part of the
hand--bag diagram as shown in fig.11. Different models of shadowing correspond
to different structure details of the diagrams of fig.11 (see
\cite{ARNEODOSHAD}).

\medskip
Shadowing is expected to be a low $x$ phenomenon. This can be understood
within the simple space--time picture of the interaction of the virtual photon
with atomic nucleus, \cite{basqcd}. In the infinite momentum frame i.e. in
a frame where the momentum $p=p_A/A$
is very large
  the wee partons (sea quarks and gluons) occupy longitudinal distances
\begin{equation}
\Delta z_p \approx {1\over xp}
\end{equation}
The momentum $p_A$ is the momentum of the nucleus
with A nucleons.
The nucleus in this frame occupies the Lorentz contracted distance:
\begin{equation}
\Delta z_A \approx 2R_A{M\over p}
\end{equation}
where $R_A$ is the nuclear radius and M the nucleon mass.  The (Lorentz
contracted) average distance between the nucleons in this frame is:
\begin{equation}
\Delta z = r{M\over p}
\end{equation}
where $r$ denotes the average distance between nucleons in the laboratory
frame.
One can distinguish three regions in $x$, i.e.:
\begin{equation}
(i) \centerline{$x>${$1\over Mr$}}
\end{equation}
that corresponds  to the partonic size being smaller than the average distance
between nucleons within nuclei:
\begin{equation}
\Delta z_p<\Delta z.
\end{equation}
In this region the shadowing is expected to be negligible since partons can be
regarded as belonging to individual nucleons.
\begin{equation}
(ii) \centerline{{$1\over 2MR_A$}$<x<${$1\over Mr$}}
\end{equation}
that corresponds to the longitudinal size of the partons being larger than
avarge distance between nucleons in a nucleus yet smaller than the (contracted)
longitudinal size of the nucleus:
\begin{equation}
\Delta z_A > \Delta z_p >\Delta z.
\end{equation}
In this region the shadowing gradually sets on with the decreasing $x$.
\begin{equation}
(iii) \centerline{$x<${$1\over 2MR_A$}}
\end{equation}
that corresponds to
\begin{equation}
\Delta z_p>\Delta z_A
\end{equation}
where shadowing is expected to be maximal.  In the regions ($i$) and ($ii$)
the partons can no longer be regarded as belonging to individual nucleons
since their longitudinal distances are larger than the average distances
between nucleons exceeding eventually the size of the nucleus.

\medskip
Let us also notice that when
the deep inelastic scattering is considered in
the laboratory frame then the three regions of $x$, i.e. $(i) - (iii)$
are defined through the
mutual relations between the lifetime $\tau_{\gamma}$ of the $q\bar q$
fluctuation of the virtual photon:
\begin{equation}
\tau_{\gamma} = {1\over Mx}
\end{equation}
and the characteristic distances $r$ and $R_A$.
The nuclear shadowing indeed is a typical small $x$ phenomenon
since $1/(Mr) \approx 0.1$.

\medskip
There exist several models of shadowing incorporating the VMD and/or
the partonic mechanisms. For a complete review of these models and a
comprehensive list of references see ref.
\cite{ARNEODOSHAD}. Since that article has been completed new papers
concerning the shadowing in the deuteron have appeared \cite{BBJKZ,MT,BARONE},
several of them inspired by the new
measurements of the $F_2^d/F_2^p$ ratio (cf. sec.8).
Understanding of shadowing effects in the deuteron has become relevant
in view of the increased precision of the measurements of the $F_2^A/F_2^d$
and $F_2^d/F_2^p$ ratios. In particular the shadowing affects extraction of
the neutron structure
function from the data. This in turn affects (decreases) the magnitude
of the experimentally evaluated Gottfried sum (\cite{zoller,BBJKZ}).

\medskip\medskip\medskip
\section{Experimental data}
The experimental problems connected to measuring and analysing the low $x$, low
$Q^2$ data  are discussed in \cite{BCKK}.  In that reference we have
also  listed the experiments which provided the then available experimental
results.
Here we limit ourselves only to updating this information.

\medskip
In the fixed target experiments the low $x$ region is correlated
with the low $Q^2$ values, see e.g. fig.12, \cite{paic}. The lowest values of
$x$ were reached by the NMC at CERN and E665 Collaboration at Fermilab through
applying special experimental techniques permitting
measurements of muon scattering angles as low as 1 mrad,
\cite{NMCR3,NMCR4,paic,E665F2}. These "small $x$
triggers" and special off-line selection methods were also effective against
a background of muons scattered
elastically from target atomic electrons; corresponding peak occurs at
$x=$0.000545. Systematic errors in both experiments, in particular
these on the
ratio of structure functions for different nuclei, $F_2^a/F_2^b$, were greatly
reduced as a result of irradiating several target materials at a time and/or of
a
frequent exchange of targets in the beam.

\medskip
During the last three years there appeared new measurements of the
$F_2^p(x,Q^2)$, $F_2^d(x,Q^2)$ \cite{NMCF2} and $F_2^d(x,Q^2)/F_2^p(x,Q^2)$
\cite{NMCR,NMCR2,NMCR3,NMCR4} by the NMC and by the E665
\cite{E665R,E665F2}. The NMC also performed the QCD analysis of their $F_2$
data
\cite{nmcqcd}. Both collaborations
have presented new and precise results on $x$, $A$ and $Q^2$ dependence
of nuclear shadowing \cite{NMCR3,paic,E665SHAD,carroll}.
The NMC also determined differences $R^{Ca}-R^C$, \cite{NMCRCa} and
$R^d-R^p$, \cite{NMCRd} at low $x$.

\medskip
Extraction of $F_2(x,Q^2)$ from the data needs information on $R(x,Q^2)$. In
particular the ratio of inelastic cross sections on different nuclei is equal
to the corresponding structure functions ratio, provided $R(x,Q^2)$ is the
same for these nuclei. Results of the NMC analysis of $R^{Ca}-R^C$ and
$R^d-R^p$ show that neither of these quantities exhibit a significant
dependence
on $x$ and that they are both compatible with zero, fig.13. In their analyses
the NMC and E665 assumed $R$ independent of the target atomic mass $A$.

\medskip
The NMC results for the deuteron structure function, $F_2^p$  as a function of
$Q^2$ for different bins of $x$, are shown in fig.14. Measurements cover ranges
0.006 $\leq x \leq$ 0.6 and 0.5 $\leq Q^2 \leq$ 55 GeV$^2$ and were the first
precise measurements at such low values of $x$. The data had great impact on
the parton distribution analysis (see e.g.\cite{MRS}) and join well to the
results of HERA, \cite{H1M}.
A clear scaling
violation pattern with slopes $d~ln F_2/d~ln Q^2$ positive at low $x$ and
negative at higher $x$ and an "approach to scaling" (i.e. $Q^2$ dependence
of $F_2$ at $Q^2$  less than few GeV$^2$) are visible in fig.14. In this figure
comparison of the NMC \cite{NMCF2}, SLAC \cite{SLAC} and BCDMS
\cite{BCDMS} measurements is also shown. The agreement between all three data
sets is very good.  At the same time the low $x$ results of the EMC NA2
\cite{EMCNA2}
experiment have been disproved by the NMC measurements.  The $x$ dependence
of the deuteron structure function at low $Q^2$ is shown in fig.15 for several
values of $Q^2$. The NMC measurements are compared  with those of EMC NA28
\cite{EMCNA28} and SLAC \cite{SLAC}.  Characteristic is a weak $x$ dependence
of $F_2^d$ at low $x$.

\medskip
The $Q^2$ dependence of the structure functions $F_2^p$ and $F_2^d$ measured by
the NMC with good accuracy down to low values of $x$ has been compared with the
predictions of perturbative QCD \cite{nmcqcd}.  The flavour singlet and
non--singlet quark distributions as well as the gluon distribution have been
parametrised at the reference scale equal to 7 GeV$^2$.  All the data with $Q^2
\geq $ 1 GeV$^2$ were included in the fit.  Besides the leading twist
contribution the higher twist term was also included in an approximate way
given by formula (\ref{F2ht}) where $H(x)$ was determined from the SLAC and the
BCDMS measurements \cite{HX}, averaged over the proton and
deuteron and suitably extrapolated to lower values of $x$.  Results of the QCD
fit to the proton structure function data are shown in fig. 16. Important here
is the extension of the QCD analysis to the low $x$ and low $Q^2$ regions.
The contribution of higher
twists is still moderate at scales about 1 GeV$^2$.

\medskip
The new (preliminary) measurements of the proton and deuteron structure
functions for $x > $0.0001 have recently been presented by the E665
Collaboration, fig.17 \cite{E665F2}. The lowest $Q^2$ values
in their data reaches a very small value of 0.2 GeV$^2$.

\medskip
Both NMC and E665 experiments have measured the deuteron to proton structure
function ratio, $F_2^d/F_2^p$, extending down to very low values of $x$.
In case of the NMC the ratio has been measured directly, i.e. the measurement
of the absolute structure function is used only for the radiative corrections
calculations.
The data are usually presented as the ratio $F_2^n/F_2^p$ where $F_2^n$ is
defined as $2F_2^d-F_2^p$. This quantity would give the structure function of
the free nucleon in the absence of nuclear effects in the deuteron. NMC results
for $F_2^n/F_2^p$ as the function of $x$
are shown in fig. 18, \cite{NMCR4}; E665 results are presented in fig. 19,
\cite{E665F2}. In the latter case instead of
the ratio $F_2^n/F_2^p$ the ratio $\sigma_n/\sigma_p$ is given.  It is equal to
the former due to $A-$independence of $R$. In both data sets the average $Q^2$
varies from bin to bin reaching down to $<Q^2>=$ 0.2 GeV$^2$ at $x=$0.0008 for
the NMC and $<Q^2>=$ 0.004 GeV$^2$ at $x=$ 5$\times$ 10$^{-6}$ for E665.
The results of both experiments show that the $F_2^n/F_2^p$ stays always
below unity down to the smallest measured values of $x$ which at low $x$
can be attributed to the nuclear shadowing in the deuteron
\cite{BBJKZ,MT,BARONE}.

\medskip
New data appeared on the nuclear shadowing. Preliminary results of a high
precision study of the $A$ dependence of nuclear shadowing by the NMC
performed in the range 0.01 $< x <$ 0.7 and 2 GeV$^2 < Q^2 <$ 60 GeV$^2$,
are shown in figures 20--22. The structure function ratios $F_2^A/F_2^C$ for
A = Be, Al, Fe and Sn together with earlier data
of SLAC \cite{SLACOLD}, show a detailed pattern of the $x$ dependence
of shadowing, fig.20. The NMC data cover the $A$ range from $A=2$ to $A=118$.
In fig.21 they are shown as a function of $A$ for three bins of $x$. The
functional dependence of $F_2^A/F_2^C$ on $A$ is approximately logarithmic
(except, possibly, the light nuclei) and the structure function ratio has
therefore been parametrised as
$F_2^A/F_2^C=a+blnA$ in each bin of $x$. The slopes, $b$, are displayed
in fig.22. The amount of shadowing increases strongly with the mass number $A$.
Much lower values of $x$ and $Q^2$ are covered by the nuclear data
obtained by the E665: $x >$0.0001 and $Q^2 >$0.1 GeV$^2$ thanks to a
special trigger and off--line analyses, \cite{carroll}, fig.23.
Shadowing seems to saturate at $x$ about 0.004 as was observed earlier
by the E665 \cite{E665SHAD} and indicated by the preliminary NMC data on
the $F_2^{Li}/F_2^D$ and $F_2^{C}/F_2^D$ ratios measured down to $x=$0.0001
and $Q^2$=0.03 GeV$^2$, \cite{paic}.
No clear $Q^2$ dependence is visible in the E665 data in a wide interval of
$Q^2$, fig.24.

\medskip\medskip\medskip
\section{Conclusions and outlook}

In this review paper we have summarised the present understanding of the
electroproduction structure functions in the region of low values of $x$ and
$Q^2$, which have recently been measured in the fixed target, charged lepton
inelastic scattering experiments.
This has included a survey of theoretical constraints and expectations,
clarification of certain definitions and concepts,
collection of the existing parametrisations of structure functions in this
region and presentation of the experimental data.
We consider this paper as being an
extension of the comprehensive review article on the small $x$ physics
\cite{BCKK} towards the more detailed treatment of the low $Q^2$ problems.
This means in particular that in the summary of the experimental data we have
presented only the results which appeared after the above mentioned article
had been completed.

\medskip
Important property of the structure function $F_2(x,Q^2)$ which
follows from the conservation of the electromagnetic current is
its linear vanishing as the function of $Q^2$ for $Q^2
\rightarrow 0$ (for fixed $\nu$).  This means that Bjorken scaling
cannot be a valid concept at low $Q^2$ and so the pure partonic
description of inelastic lepton scattering has to break down
for moderate and low $Q^2$ values. At moderately large $Q^2$ the higher twist
contributions to $F_2$ which vanish as negative powers of $Q^2$ are
frequently being included in the QCD data analysis. One also expects that at
low
$Q^2$ the VMD mechanism should play an important role.

\medskip
The small $x$ behaviour of the structure function
$F_2(x,Q^2)$ is dominated by the
pomeron exchange. Analysis of the structure function in the
small $x$ region for both low and moderate values of $Q^2$
can clarify our understanding of the pomeron, i.e. possible
interplay between the "soft" and "hard" pomerons, the role of
shadowing (or absorptive) corrections, etc. At
 large $Q^2$ the problem is linked  with the QCD expectations
concerning the deep inelastic scattering at small $x$, see
ref.\cite{SMX} for a recent review.
Besides the structure functions (or total cross sections)
complementary information on the pomeron can also be obtained
from the analysis of diffractive processes in the electro- and
photoproduction. This concerns both the inclusive diffractive
processes and the diffractive production of vector
mesons; the present experimental situation is described in refs.
\cite{H1M,ZEUSM,WOLFSA,EICHLER}.

\medskip
Descriptions of the low $Q^2$, low $x$ behaviour of $F_2$ range from pure fits
to experimental data to the dynamically motivated models.
The existing parametrisations have been collected  in a way which should make
it easy to use in practical applications,
e.g. in the radiative correction procedure. Wherever possible a dynamical
content of each parametrisation was exposed.

\medskip
Since 1992 a wealth of measurements of $F_2$ has been published. In the region
of interest for this paper this included the NMC and E665 results extending
down to very low $x$ and $Q^2$ values and displaying characteristic "approach
to scaling" behaviour. Nuclear shadowing phenomenon was studied in great
detail for targets ranging from $A=2$ to $A=118$ by the same two
collaborations.
Its $x$, $Q^2$ and $A$ dependence was precisely measured.

\medskip
The new possibilities have opened up with the advent of the HERA collider.
Presently the data there are collected at $Q^2$ larger than approximately
5 GeV$^2$ and therefore have not been discussed in this review. These data
show a very strong increase of $F_2$ with decreasing $x$, \cite{H1,ZEUS}.
Let us remind that
for fixed $Q^2$ the increase of the structure function $F_2(x,Q^2)$
with decreasing $x$ reflects the increase of the virtual Compton scattering
total cross section with $W^2$.
We would like to emphasise that this increase observed at large $Q^2$ is much
stronger than the increase of the total photoproduction cross section
observed between the fixed target and the HERA energies. Dynamical
understanding
of this evident variation of the $W^2$ dependence with $Q^2$ is certainly
very interesting and important.  The experimental data at low values of $Q^2$
($Q^2 \simeq 1$ GeV$^2$) which would cover similar range of $W^2$ as those
at high $Q^2$ would be extremely valuable for this purpose.  It would therefore
be very interesting to extend the HERA measurement down to the low values
of $Q^2$ as it has been planned recently \cite{ADR}.

\medskip\medskip\medskip
\section{Acknowledgments}
We thank our colleagues from the NMC for numerous discussions and
from the E665 collaboration for providing us with their data.
This research has been supported in part by the Polish State Committee for
Scientific Research grant number 2 P302 062 04.

\medskip\medskip\medskip
\noindent
{\Large {\bf Figure Captions}}
\medskip\medskip
\begin{enumerate}
\item
Illustration of the continuity of physical processes: double differential
cross section for the electron--proton inelastic scattering is sketched as a
function of the energy transfer $\nu$ for different values of the resolution
$Q^2$. Dashed and continuous lines correspond to constant values of $x$ and
$W$ respectively. Definitions of kinematical variables are given in sec.2.
\item
Kinematics of inelastic charged lepton--proton scattering in the one photon
exchange approximation and its relation through the optical theorem to Compton
scattering for the virtual photon.
\item
Total photoproduction cross section as a function of the $\gamma p$ centre of
mass energy, $W_{\gamma p}$. Low energy data come from ref.\cite{compil},
recent
ZEUS Collaboration data -- from ref.\cite{HERAPHOTnew}, previous data -- from
ref.\cite{HERAPHOT}. The curves are explained in the text. Figure comes from
ref.\cite{HERAPHOTnew}.
\item
Parametrisation of $F_2^p(x,Q^2)$ by Donnachie and Landshoff as a function of
$Q^2$ for the following values of $x$ (from the top): 0.008, 0.035, 0.07, 0.18
and 0.45 together with data from NMC \cite{NMCF2}, SLAC \cite{SLAC} and
BCDMS \cite{BCDMS}. The $F_2$ values are scaled by factors (from the top):
16, 8, 4, 1.5 and 1 for clarity. The error bars represent
statistical and systematic errors added in quadrature (from \cite{radcor1}).
\item
Parametrisation of $F_2^p(x,Q^2)$ by Abramowicz et al. The $x$ values, data
and scaling factors are as in fig.4 (from \cite{radcor1}).
\item
Parametrisation of $F_2^p(x,Q^2)$ by Bade{\l}ek and Kwieci\'nski as a function
of $Q^2$ for the following values of $x$ (from the top): 0.008, 0.0125, 0.0175,
0.035. 0.07 and 0.1 together with data from NMC \cite{NMCF2}, SLAC \cite{SLAC}
and BCDMS \cite{BCDMS}. The $F_2$ values are scaled by factors
(from the top): 32, 26, 8, 4, 2 and 1 for clarity.
The error bars represent statistical and systematic errors added in quadrature
(from \cite{radcor1}).
\item
Regions of the ($x,\mu^2$) plane as used in the parametrisation by Schuler and
Sj\"ostrand (from \cite{ss}).
\item
Comparison of the parametrisations of $F_2^p(x,Q^2)$ by Abramowicz et al. (full
curves), Bade{\l}ek and Kwieci\'nski (dashed curves) and Donnachie and
Landshoff
(dash--dotted curves) as functions of $Q^2$ for the following values of $x$
(from the top): 0.008, 0.02, 0.05 and 0.1. The values are scaled by the factors
(from the top): 27, 9, 3 and 1 for clarity (from \cite{radcor1}).
\item
Sketch of the $x$ dependence of the ratio $F_2^A/F_2^d$ where various nuclear
effects are indicated.
\item
Multiple scattering of vector mesons. Lines in the lower part of the diagram
denote nucleons.
\item
Hand--bag diagram for the virtual Compton scattering on a nucleus and its
decomposition into a multiple scattering series. Lines in the upper parts
of diagrams denote quarks (antiquarks) and lines in the lower
part denote nucleons.
\item
Kinematic ranges covered by different triggers in the NMC (from \cite{paic}).
\item
NMC (preliminary) results $R^d-R^p$ as a function of $x$ compared with the
QCD predictions (the curve, see \cite{NMCRd} for details) and with the results
of SLAC (open symbols, \cite{SLACR}). Figure comes from ref. \cite{NMCRd}.
\item
The $F_2^d$ data from NMC compared to the data from SLAC \cite{SLAC} and
BCDMS \cite{BCDMS}. The error
bars are the quadratic sums of the statistical and systematic errors, excluding
the overall normalisation uncertainty.  The curve is a result of a 15-parameter
function fit to all three data sets, given in Table 1 (from \cite{NMCF2}).
\item
Comparison of the (low $Q^2$) $x$ dependence of the NMC \cite{NMCF2}, EMC NA28
\cite{EMCNA28} and SLAC \cite{SLAC} results (from \cite{NMCF2}).
\item
Results of the QCD fit to the $F_2^p$ data. The solid line is the result
of the QCD fit with higher twist included, cf. eq. (\ref{F2ht}).
The dotted curve shows the contribution of $F_2^{LT}$. In
the fit 90 (280) GeV data were renormalised by 0.993 (1.011).  The errors are
statistical (from \cite{nmcqcd}).
\item
Preliminary measurements of $F_2^p(x,Q^2)$ by the E665 Collaboration at
Fermilab. The errors are statistical. Systematic errors
are 5--15 $\%$ (from \cite{E665F2}).
\item
NMC results on the ratio $F_2^n/F_2^p$ as the function of $x$ at $Q^2$ values
averaged in each $x$ bin.  The solid symbols mark the final data set while the
open ones are still perliminary data taken with a special "small $x$
trigger".  Errors are statistical; the band at the bottom indicates the
preliminary estimate of systematic errors (from \cite{NMCR4}).
\item
E665 results for $\sigma_n/\sigma_p$ as a function of $x$ at $Q^2$ values
averaged in each $x$ bin for three methods of data analysis.
Errors are statistical; the total systematic uncertainty is less than 3.5$\%$
in all $x$ regions.   In this figure
the NMC data at $Q^2= $4 GeV$^2$ are also shown (from \cite{E665F2}).
\item
NMC results on structure function ratios, $F_2^A/F_2^C$, as functions of $x$
together with the earlier results of SLAC \cite{SLACOLD} (from \cite{NMCR3}).
Errors are statistical.
\item
$A$ dependence of the NMC $F_2^A/F_2^C$ data for three $x$ bins (from
\cite{NMCR3}).
\item
Slopes $b$ from the $F_2^A/F_2^C=a+blnA$ fit to the NMC data
(from \cite{NMCR3}).
\item
E665 results on $\sigma^A/\sigma^D$ as a function of $x$.
Errors are statistical, the systematic ones are marked as shaded bands.
The 3.4$\%$ overall normalisation error has not been included
(from \cite{carroll}).
\item
$Q^2$ dependence of the E665 nuclear data in bins of $x$ (from \cite{carroll}).
\end{enumerate}

\end{document}